\documentclass{article}
\usepackage{sita2007}
\usepackage{color}
\usepackage{amsmath}
\usepackage{amssymb}
\usepackage{graphicx}

\title{
Message-Passing Decoding of Lattices Using Gaussian Mixtures
}

\author{
Brian M. Kurkoski%
\thanks{Dept. of Information and Communications Engineering, Univ. of Electro-Communications, Tokyo, Japan, kurkoski@ice.uec.ac.jp.}
  \and
Justin Dauwels\thanks{Amari Research Unit,
RIKEN Brain Science Institute,
Saitama, Japan, justin@dauwels.com.}
}

\newcommand{\Nbb}[3]{\mathcal{N}\!\left({#1};{#2},{#3}\right)}

\begin{document}
 
\abstract{
A lattice decoder which represents messages explicitly as a mixture of Gaussians functions is given.  In order to prevent the number of functions in a mixture from growing as the decoder iterations progress, a method for replacing $N$ Gaussian functions with $M$ Gaussian functions, with $M \leq N$, is given.  A squared distance metric is used to select functions for combining.  A pair of selected Gaussians is replaced by a single Gaussian with the same first and second moments.  The metric can be computed efficiently, and at the same time, the proposed algorithm empirically gives good results, for example, a dimension 100 lattice has a loss of 0.2 dB in signal-to-noise ratio at a probability of symbol error of $10^{-5}$.
}


\maketitle

\tolerance=5000

\section{Introduction}

Lattices play a central role in many communication problems.  While Shannon used a non-lattice, and non-constructive, Euclidean-space code to compute the capacity of the AWGN channel, recently Erez and Zamir showed that lattice encoding and decoding can also achieve the capacity of the AWGN channel \cite{Erez-it04}.  Similarly, for the problem of communication with known noise, which has applications to multiuser communications and information hiding, lattice codes play an important role \cite{Erez-it05*2}.  In source coding, lattices may be used for lossy compression of a real-valued source.

To approach theoretical capacities, it is necessary to let the dimension of the lattice or code become asymptotically large.  However, for most lattices of interest, the decoder complexity is worse than linear in the dimension, and most studied lattices have small dimension.  For example, a frequently cited reference on lattice decoding gives experimental results with a maximum dimension of 45 \cite{Agrell-it02}.  Other approaches use trellis-based lattices, which are exponentially complex in the number of states \cite{Erez-it05*3}.  Historically, finite-field error correcting codes also suffered the same complexity limitation, however, with the advent of iteratively-decoded low-density parity check codes and turbo codes, the theoretical capacity of some binary-input communication channels can be achieved \cite{Richardson-it01*2}.

Recently, a new lattice construction and decoding algorithm, based upon the ideas of low-density parity check codes has been introduced.  So-called low-density lattice codes (LDLC) are lattices defined by sparse inverse generator matrix with a pseudo-random construction.  Decoding is performed iteratively using message-passing, and complexity is linear in the block length.  Sommer, Feder and Shalvi, who proposed this lattice and decoder, demonstrated decoding with dimension as high as $10^6$.  However, the experiments considered decoding only for a special communications problem where the transmit power is unconstrained.  Comments in their paper suggest that the algorithm did not converge when applied to the more important problem of general lattice decoding  \cite{Sommer-isit06} \cite{Sommer-itsub}.

When decoding on the AWGN channel, the LDLC decoder messages are continuous-valued functions, which can be exactly represented by a mixture of Gaussian functions.   However, as iterations progress, the number of Gaussians in the mixture grows rapidly.  A direct implementation of a decoder which exploits this property is infeasible, and so prior works quantize the messages, ignoring the Gaussian nature of the messages.  

In this paper, the LDLC decoder messages are represented as Gaussian functions, and the growth in the number of Gaussians is reduced by a proposed Gaussian mixture reduction algorithm.  This algorithm approximates a number of Gaussians $N$ with a smaller number of Gaussians $M$.  The algorithm combines Gaussians in a pair-wise fashion iteratively until a stopping condition is reached.  A distance metric, which computes the squared difference between a pair of Gaussian functions, and the single Gaussian which has the same first and second moments, is used.

Section \ref{sec:ldlc} gives a review of the construction and decoding algorithm for low-density lattice codes. If the channel noise is Gaussian, then messages in the decoding algorithm can be represented as a mixture of Gaussian functions.  Section \ref{sec:gm} gives a method for replacing a pair of Gaussians with a single Gaussian, which is applied to an algorithm which reduces a mixture of $N$ Gaussian functions to a mixture of $M$ Gaussians.  Section \ref{sec:apply} applies this algorithm to the decoding of low-density lattice codes, and considers simulation results.  Section \ref{sec:conclusion} is the conclusion.

\section{Low-density Lattice Codes \label{sec:ldlc}}

\subsection{Lattices and Lattice Communication}

A lattice is a regular infinite array of points in $\mathbb R^n$.

\emph{Definition} An $n$-dimensional lattice $\Lambda$ is the set of points $\mathbf x = (x_1,x_2, \ldots, x_n)$ with
\begin{eqnarray}
\mathbf x &=& G \mathbf b, \label{eqn:lattice}
\end{eqnarray}
where $G$ is an $n$-by-$n$ generator matrix and $\mathbf b =(b_1,\ldots, b_n)$ is the set of all possible integer vectors, $b_i \in \mathbb Z$.

The following communications system is considered.  Let the codeword $\mathbf x$ be an arbitrary point of the lattice $\Lambda$.  This codeword is transmitted over an AWGN channel with known noise variance $\sigma^2$, and received as the sequence $\mathbf y = \{y_1, y_2, \ldots, y_n\}$:
\begin{eqnarray}
y_i &=& x_i + z_i, i=1,2,\ldots,n,
\end{eqnarray}
where $z_i$ is the AWGN.  A maximum-likelihood decoder selects $\widehat{\mathbf x}$ as the estimated codeword:
\begin{eqnarray}
\widehat {\mathbf x} &=& \arg \max_{\mathbf x \in \Lambda} Pr( \mathbf y | \mathbf x)
\end{eqnarray}

The received codeword is correct if $\mathbf x = \widehat{ \mathbf x}$ and incorrect otherwise.  The power of the transmitted symbol, $|| \mathbf x||^2$ is unbounded.  Instead, power is restricted by the volume of the Voronoi region, $\det(G)$.

For this system, Poltyrev \cite{Poltyrev-it94} showed that for sufficiently large $n$, there exists a lattice for which the probability of error becomes arbitrarily small, if and only if,
\begin{eqnarray}
\sigma^2 &<& \frac{|\det(G)|^{2/n}}{2 \pi e}.
\end{eqnarray}

Poltyrev's result is in contrast to Shannon's theorem that the capacity of the Gaussian channel, subject to a transmission power constraint, is $\frac{1}{2} \log (1+ \textrm{SNR})$.  To achieve capacity while observing the power constraint, the codepoints are on the surface of an $n$-sphere with high probability.

\subsection{LDLC Definition}

\emph{Definition} A low-density lattice code is a lattice with a non-singular generator matrix $G$, for which $H =  G^{-1}$ is sparse.

Regular LDLC's have $H$ matrices with constant row and column weight $d$.  Although not necessary, it is convenient to assume that  $\det(H) = 1/\det(G) = 1$.  The non-zero entries are selected pseudo-randomly.

In a \emph{magic square LDLC}, the absolute values of the $d$ non-zero entries in each row and each column are drawn from the set $\{h_1, h_2, \ldots, h_d\}$ with  $h_1 \geq h_2 \geq \cdots \geq h_d > 0$.  The signs of the entries of $H$ are pseudo-randomly changed to minus with probability 0.5.    From here, $(n,d)$ magic square LDLC's are considered with $h_1 = 1$, and $h_i = 1/\sqrt{d}$ for $i=2,\ldots, d$. Such codes resulted in only slightly worse performance than other weight sequences \cite{Sommer-itsub}.

\subsection{LDLC Decoding}

The LDLC decoding algorithm is based upon belief-propagation, where messages are real functions corresponding to probability distributions on the symbols $x_i$.  As with decoding low-density parity check codes, the decoding algorithm may be presented on a bipartite graph.  There are $nd$ variable-to-check messages $q_k(z)$, and $nd$ check-to-variable messages $r_k(z)$,  $k=1,2,\ldots, nd$.

With an AWGN channel, the initial message is:
\begin{eqnarray}
q_k(z) &=& \frac{1}{\sqrt{ 2 \pi } \sigma} e^{ - \frac{ (y_i - z)^2}{2 \sigma^2}},  \label{eqn:channelmessage}
\end{eqnarray}
for the edge $k$ connected to variable node $i$.

\subsubsection{Check Node}

For the check node, note that (\ref{eqn:lattice}) can be re-written as:
\begin{eqnarray}
H \mathbf x &=& \mathbf b,
\end{eqnarray}
which defines a sparse system of equations:
\begin{eqnarray}
h_{i j_1} x_{j_1} + h_{i j_2} x_{j_2} + \cdots + h_{i j_1} x_{j_1} &=& b_i, \label{eqn:sseq}
\end{eqnarray}
for $i = 1, 2, \ldots, n$, and $j_k \in \mathcal I_i$, where $\mathcal I_i$ is the columns of $H$ which have a non-zero entry in position $i$.

Let $\widetilde x_k = h_k x_k$, so $\sum_{i=1}^d \widetilde x_i = b$, where $b$ is an integer.  The input and output messages are $q_k(z)$ and $r_k(z)$, respectively, for $k=1,2,\ldots, d$.  From (\ref{eqn:sseq}), for an arbitrary $i$,  $x_k=$
\begin{eqnarray}
 \frac{ b - (h_1 x_1 + \cdots +h_{k-1} x_{k-1} + h_{k+1} x_{k-1} + \cdots + h_d x_d) }{h_k} \nonumber,
\end{eqnarray}
or, 
\begin{eqnarray}
x_k & = &\frac{1}{h_k} ( b - \sum_{i=1}^{d \setminus k} \widetilde x_i) \label{eqn:simplecbp}.
\end{eqnarray}
The output message $r_k(z)$ can be obtained from the input messages $q_i(z), i=1,\ldots,d, i \neq k$ in four steps, Unstretch, Convolution, Extension and Stretch.

\emph{Unstretch} is multiplication by $h_k$.  The message for $\widetilde x_i$ is $\widetilde q_k(z)$, 
\begin{eqnarray}
\widetilde q_k(z) &=& q_k( \frac{z}{h_k}).
\end{eqnarray}

\emph{Convolution} The message for $\sum_{i=1}^{d \setminus k} \widetilde x_i$ is $\widetilde r_k(z)$.  The distribution of the sum of random variables is the convolution of distributions,
\begin{eqnarray}
\widetilde r_k(z) &=& (\widetilde q_1 * \cdots * \widetilde q_{k-1}  * \widetilde q_{k+1} * \cdots * \widetilde q_{d})(z) , \label{eqn:convolution}
\end{eqnarray}
where $*$ denotes real-number convolution.

\emph{Extension} is a shift-and-repeat operation for the unknown integer $b$.  Conditioned on a specific value of $b$, the distribution of $b - \sum_{i=1}^{d \setminus k} \widetilde x_i$ is $\widetilde r_k(b-z)$.    Assuming that $b$ is an arbitrary integer with uniform a priori distribution, 
\begin{eqnarray}
\widetilde r'_k(z) &=& \sum_{b=-\infty}^{\infty} \widetilde r_k(b-z)  \label{eqn:extension}.
\end{eqnarray}

\emph{Stretching} is multiplication by $1/h_k$.  Finally the message $r_k(z)$ which is the message for (\ref{eqn:simplecbp}), is obtained as:
\begin{eqnarray}
r_k(z) &=& \widetilde r'_k( h_k z)  \label{eqn:stretching}
\end{eqnarray}

Note that the above operations are linear and can be interchanged as is required for an implementation.

\subsubsection{Variable Node}

At variable node $i$, take the product of incoming messages, and normalize.

\emph{Product:}
\begin{eqnarray}
\widehat q_k(z) &=& e^{- \frac{(y_i - z)^2}{2 \sigma^2}} \prod_{i=1}^{d \setminus k} r_i(z). \label{eqn:varnode}
\end{eqnarray}

\emph{Normalize:}
\begin{eqnarray}
q_k(z) &=& \frac{ \widehat q_k(z) }{\int_{-\infty}^{\infty} \widehat q_k(z) dz} .
\end{eqnarray}

\subsubsection{Estimated Codeword and Integer Sequence}

The check node and variable node operations are repeated iteratively until a stopping condition is reached.   Estimate the transmitted by codeword $\widehat{\mathbf x}$ by first computing the a posteriori message $F_i(z)$ for the code symbol $x_i$ as:
\begin{eqnarray}
F_i(z) &=& e^{- \frac{(y_i - z)^2}{2 \sigma^2}} \prod_{k=1}^{d} r_k(z).
\end{eqnarray}
Find $\widehat x_i$ as:
\begin{eqnarray}
\widehat x_i &=& \arg \max_{z \in \mathbb{R}}  F_i(z).
\end{eqnarray}
The estimated integer sequence $\widehat{\mathbf b}$ is:
\begin{eqnarray}
\widehat{\mathbf b} &=& \langle H \widehat{\mathbf x} \rangle,
\end{eqnarray}
where $\langle z \rangle$ denotes the integer closest to $z$.
 
\subsection{Gaussian Mixture Decoder \label{subsec:gmd}}

When the channel noise is Gaussian, all of the LDLC messages can be described as a mixture of Gaussian functions.  From here, ``Gaussians'' will be used as shorthand for ``Gaussian functions''.

In this section, it is assumed that a message $f(z)$ is a mixture of $N$ Gaussians,
\begin{eqnarray}
f(z) = \sum_{i=1}^{N} c_i \Nbb{z}{m_i}{v_i}, \label{eqn:mixture}
\end{eqnarray}
where $c_i \geq 0$ are the mixing coefficients with $\sum_{i=1}^{N} c_i = 1$, and 
\begin{eqnarray}
\Nbb z m v = \frac{1}{\sqrt{2 \pi v}} e^{ - \frac{ (z-m)^2}{2 v}}.
\end{eqnarray}
In this way, the message $f(z)$ can be described by a list of triples of means, variances and mixing coefficients, $\{(m_1, v_1, c_1), \ldots, (m_{N}, v_{N}, c_{N} ) \}$

In describing the Gaussian mixture decoder, initially assume that the input messages to a node consist of a single Gaussian, that is $N = 1$.

\emph{Check node} Without loss of generality, consider check node inputs $k = 1,2,\ldots,d-1$ and output $d$.  Each input message $q_k(z)$ is a single Gaussian $\Nbb{z}{m_k}{v_k}$.

The message $\widetilde q_k(z)$ is obtained by multiplying by $h_k$, so $\widetilde q_k(z) = \Nbb z {h_k m_k}{h_k^2 v_k}$.

The message $\widetilde r_d(z)$ is the convolution of $\widetilde q_k(z), k=1,\ldots,d-1$.  So:
\begin{eqnarray}
\widetilde r_d(z) & = & \Nbb z {\sum_{k=1}^{d-1} h_k m_k}{ \sum_{i=1}^{d-1}  h_k^2 v_k}.
\end{eqnarray}

The message $\widetilde r'_d(z)$ is  $\widetilde r_d(z)$ shifted over all possible integers:
\begin{eqnarray}
\widetilde r'_d(z)\! \! \!  & = &\! \!  \! \sum_{b=-\infty}^{\infty} \! \! \Nbb z {\sum_{k=1}^{d-1} h_k m_k + b}{ \sum_{k=1}^{d-1}  h_k^2 v_k}. \nonumber 
\end{eqnarray}

The output message $r_d(z)$ is obtained by scaling by $-1/ h_d$, so:
\begin{eqnarray}
r_d(z) \! \! \! & = & \! \! \!  \sum_{b=-\infty}^\infty \Nbb z {- \frac{\sum_{k=1}^{d-1} h_k m_k + b}{h_d}} { \frac{\sum_{k=1}^{d-1}  h_k^2 v_k }{h_d^2} } . \nonumber 
\end{eqnarray}

\emph{Variable Node}.  Let the check-to-variable node messages $r_k(z), k = 1,\ldots, d-1$ be Gaussians $\Nbb z {m_k} {v_k}$.  For notational convenience, let $m_0 = y_i$ be the symbol received from the channel at node $i$ and let $v_0 = \sigma^2$ be the channel variance, as in (\ref{eqn:channelmessage}).  The output message $q_d(z)$, the product of these input messages, will also be a Gaussian,
\begin{eqnarray}
q_d(z) &=& k_d \Nbb z {m_d} {v_d},
\end{eqnarray}
where,
\begin{eqnarray}
\frac{1}{v_d} &=& \sum_{k=0}^{d-1}  \frac{1}{v_k}, \\
\frac{m_d}{v_d} &=& \sum_{k=0}^{d-1} \frac{m_{k}}{v_{k}} \\[-24pt] \nonumber
\end{eqnarray}
and,
\begin{eqnarray}
k_d\! \! \! \! \!  & =  &\! \! \! \! \!  \sqrt{ \frac{v_d}{(2 \pi)^{d-2} \prod_i v_i}}  \exp \!  \Big(\! \!  - \frac{v_d}{2} \sum_{i=0}^{d-2} \sum_{j=i+1}^{d-1} \frac{ (m_i - m_j)^2}{v_i v_j} \Big ) .  \nonumber 
\end{eqnarray}

For the general case where the input consists of a mixture of Gaussians, at either the check node or the variable node, the output can be found by conditioning on one element from each input mixture and computing a single output Gaussian.  The mixing coefficient for this Gaussian is the product of the input mixing coefficients.  Then the output is the mixture of these single Gaussians created by conditioning all input combinations.

The number of Gaussians in each mixture grows rapidly as the iterations progress.  At the variable node, if input $k$ consists of a mixture of $N_k$ Gaussian functions, then the output message will consist of $N_1 N_2 \cdots N_{d-1}$ Gaussian functions.   At the check node, even if the number of integer shifts is bounded, the number of Gaussian functions in the mixture also grows as $O(N^{d-1})$.  A naive implementation of this Gaussian mixture decoder is prohibitively complex.  The following section proposes a technique for approximating a large number of Gaussians.

\section{Gaussian Mixture Reduction \label{sec:gm} }

This section describes an algorithm which approximates a mixture of Gaussian functions with a smaller number of Gaussian functions.  

The algorithm input is a mixture of $N$ Gaussians, $f(z)$, as defined in (\ref{eqn:mixture}), given as a list of triples.  The algorithm output is a list of $M$ triples of means, variances and mixing coefficients, $\{$ $(m^{\mathsf m}_1, v^{\mathsf m}_1, c^{\mathsf m}_1),$ $\ldots,$ $(m^{\mathsf m}_M, v^{\mathsf m}_M, c^{\mathsf m}_M)$ $ \} $  with $\sum_{i=1}^M c^{\mathsf m}_i  =1$, that similarly forms a Gaussian mixture $g(z)$.  With $M \leq N$, the output mixture should be a good approximation of the input mixture:
\begin{eqnarray}
f(z) &\approx& g(z) = \sum_{i=1}^M c^{\mathsf m}_i \Nbb z {m^{\mathsf m}_i} {v^{\mathsf m}_i}.
\end{eqnarray}

First, a metric which describes the error due to replacing a two Gaussians with a single Gaussian is given.  Then, this is incorporated into a greedy search algorithm which replaces $N$ Gaussians with $M$ Gaussians. 

\subsection{Approximating a Mixture of Two Gaussians with a Single Gaussian}

\emph{Definition} The squared 
difference $\textrm{SD}(p || q)$ between
 two distributions $p(z)$ and $q(z)$ with 
 support $\mathcal Z$ is defined as:
\begin{eqnarray}
\textrm{SD}(p || q) &=& \int_{z \in \mathcal Z} (p(z) - q(z))^2 dz
\end{eqnarray}
 
\emph{Lemma}  The squared difference $\textrm{SD}(p||q)$ has the following properties: 
\begin{itemize}
\item SD$(p||q) \geq 0$ for any distributions $p$ and $q$. 
\item SD$(p||q)$ if and only if $p = q$. 
\item SD$(p||q) = \textrm{SD}(q||p)$. 
\end{itemize}

\emph{Lemma} The squared difference between the Gaussian distributions $\mathcal N(m_1, v_1)$ and $\mathcal N(m_2,v_2)$ is given by $\textrm{SD}(\mathcal N(m_1,v_1), \mathcal N(m_2,v_2) )=$
\begin{eqnarray}
\frac{1}{2 \sqrt{\pi v_1} } +\frac{1}{2 \sqrt{\pi v_2} } - \frac{2}{\sqrt{2 \pi (v_1 + v_2)}} e^{ - \frac{ (m_1 - m_2)^2}{2 (s_1 + s_2)}} .
\end{eqnarray}

\emph{Lemma} The squared difference between a single Gaussian $\mathcal N(m,v)$ and a mixture of two Gaussians $c_1 N(m_1,v_1)$ $+$ $c_2 N(m_2,v_2)$, with $c_1 + c_2 = 1$, is:
\begin{eqnarray}
& & \frac{1}{2 \sqrt{\pi v}}  
+ \frac{c_1^2}{2 \sqrt{\pi v_1}} 
+ \frac{c_2^2}{2 \sqrt{\pi v_2}}  \nonumber \\
& & - \frac{2 c_1}{\sqrt{2 \pi (v+v_1)}} e^{ - \frac{ (m - m_1)^2}{2 (v+v_1)}} 
- \frac{2 c_2}{\sqrt{2 \pi (v + v_2)}} e^{ - \frac{ (m-m_2)^2}{2 (v + v_2)}} \nonumber \\ 
& & + \frac{2 c_1 c_2}{\sqrt{2 \pi (v_1 + v_2)}} e^{ - \frac{(m_1- m_2)^2}{2(v_1+v_2)} } \label{eqn:sd12}.
\end{eqnarray}

There is unfortunately no closed-form expression for the minimal
squared difference in the previous lemma.  However, minimizing the
Kullback-Leibler divergence between the single Gaussian distribution and the mixture of two Gaussian distributions is
tractable; it simply amounts to moment matching. Therefore, from now we will
consider the moment-matched Gaussian approximation.

\emph{Lemma}
The mean $m$ and variance $v$ of a mixture of two Gaussian
distributions $c_1\,\mathcal{N}(m_1,v_1)+c_2\,\mathcal{N}(m_2,v_2)$ are given by:
\begin{eqnarray}
m&=&c_1m_1+c_2m_2 \label{eqn:mean} \\
s&=&c_1 (m_1^2+v_1)+ c_2
(m_2^2+v_2) \nonumber \\ & & 
-c_1^2m_1^2-2c_1c_2m_1m_2-c_2^2m_2^2. \label{eqn:variance}
\end{eqnarray}

Let $\overline t_i, i=1,2$ denote the triple $(m_i, v_i, \overline c_i)$, where $\overline c_1 + \overline c_2$ is not necessarily one, and let the normalized triple be $t_i = (m_i, v_i, c_i /(c_1 + c_2))$.   The single Gaussian which satisfies the property of the Lemma is denoted as:
\begin{eqnarray}
t = \textrm{MM}(\overline t_1, \overline t_2),
\end{eqnarray}
where $t = (m,v,1)$, with $m$ and $v$ as given in $(\ref{eqn:mean})$ and $(\ref{eqn:variance})$.

\emph{Definition} The Gaussian quadratic loss $\textrm{GQL}(p)$ of a probability distribution
$p$ is defined as the squared difference between $p$ and the
Gaussian distribution with the same mean $m$ and variance $v$ as $p$:
\begin{equation}
\textrm{GQL}(p)=\textrm{SD}(p\|\,\mathcal{N}(m,s)).
\end{equation}

\emph{Corollary}
The Gaussian quadratic loss of a mixture of two Gaussian
distributions,
\begin{eqnarray*}
\textrm{GQL}(t_1, t_2)\! =\! \textrm{SD}( c_1 \mathcal N (m_1, v_1)\! +\! c_2 \mathcal N(m_2,v_2) \| \mathcal N(m,v) ),
\end{eqnarray*}
is obtained evaluating (\ref{eqn:sd12}), with $m$ and $v$ as given in (\ref{eqn:mean}) and (\ref{eqn:variance}).

\subsection{Approximating $N$ Gaussians with $M$ Gaussians   }

Here, we use the results from the previous subsection and propose an algorithm which approximates a mixture of $N$ Gaussians with a mixture of $N$ Gaussians.

Input: list $\mathcal L = \{t_1, t_2, \ldots, t_N\}$ of $N$ triples describing a Gaussian mixture, and two stopping parameters, $\theta$ the allowable one-step error (measured by GQL) and $M$, the maximum number of allowable Gaussians in the output.

\noindent \emph{Algorithm}
\begin{enumerate}
\item Initialize the current search list, $\mathcal C$, with the input list: $\mathcal C \leftarrow \mathcal L$.
\item Initialize the current error, $\theta^{\mathsf c}$, to the minimum GQL between all pairs of Gaussians: 
\begin{eqnarray*}
 \theta^{\mathsf c} = \min_{t_i,t_j \in \mathcal C,i \neq j} \textrm{GQL}( t_i , t_j ).
 \end{eqnarray*}
\item Initialize length of current list, $M^{\mathsf c}  = N$.
\item While $\theta^{\mathsf c} < \theta$ or $M^{\mathsf c} > M$:
\begin{enumerate}
\item  Determine the pair of Gaussians $(t_i,t_j)$ with the smallest GQL:
\begin{eqnarray*}
(t_i,t_j) &=& \arg \min_{t_i,t_j \in \mathcal C,i \neq j} \textrm{GQL}(t_i, t_j).
\end{eqnarray*}
\item Add the single Gaussian with the same moment as $t_i$ and $t_j$ to the list:
\begin{eqnarray*} 
\mathcal C \leftarrow \mathcal C \cup MM(t_i, t_j).
\end{eqnarray*}
\item Delete $t_i$ and $t_j$ from list:  $\mathcal C \leftarrow \mathcal C \setminus \{t_i, t_j\}$. 
\item Recalcuate the minimum GQL: 
\begin{eqnarray*}
 \theta^{\mathsf c} = \min_{t_i,t_j \in \mathcal C,i \neq j} \textrm{GQL}( t_i ,  t_j ).
 \end{eqnarray*}
\item Decrement the current list length:  $M^{\mathsf c}  \leftarrow M^{\mathsf c} - 1$.
\end{enumerate}
\item Algorithm output: list of triples $\mathcal C$
\end{enumerate}

Note that two conditions must be satisfied for the algorithm to stop.  That is, the one-step error may be greater than the threshold $\theta$ if the minimum number of Gaussians is not yet met.  On the other hand, the number of output Gaussians may be less than $M$, if the one-step error is sufficiently low.

\section{Gaussian-Mixture Reduction Applied to LDLC Decoding \label{sec:apply} }

In this section, the Gaussian mixture reduction algorithm of Section \ref{sec:gm} is applied to the LDLC decoding algorithm described in Section \ref{subsec:gmd}.

At the check node, observe that the message $\widetilde r_k(z)$, as given in (\ref{eqn:convolution}), can be computed recursively with $a_k(z)$ and $b_k(z)$ defined as:
\begin{eqnarray}
a_1(z) &=& \widetilde q_1(z), \\
a_k(z) &=& a_{k-1}(z) * \widetilde q_k(z), k=2 \ldots, d-1, \label{eqn:forward}
\end{eqnarray}
and, 
\begin{eqnarray}
b_d(z) &=& \widetilde q_d(z), \\
b_k(z) &=& b_{k+1}(z) * \widetilde q_k(z), k=d-1, \ldots,2.  \label{eqn:backward}
\end{eqnarray}
Then $\widetilde r_k(z)$ is found using a variation on the forward-backward algorithm as:
\begin{eqnarray}
\widetilde r_1(z) &=& b_2(z), \\
\widetilde r_k(z) &=& a_{k-1}(z) * b_{k+1}(z), \nonumber \\
& & k=2,3,\ldots, d-1 \textrm{ and,}\\
\widetilde r_d(z) &=& a_{d-1}(z).
\end{eqnarray}

The Gaussian mixture reduction algorithm is applied after the computation (\ref{eqn:forward}) and (\ref{eqn:backward}), for each $k$.  For example if $\overline a_k(z)$ is the mixture produced by applying the Gaussian mixture reduction algorithm to $a_k(z)$,
\begin{eqnarray}
\overline a_k(z) &=& \textrm{GMR}( a_k(z)),
\end{eqnarray}
then the forward recursion of the check node function may be stated as:
\begin{eqnarray}
\overline a_1(z) &=& \widetilde q_1(z) , \\
& \makebox[0cm][r]{ $\textrm{For } k=2,3,\ldots,d-1$:} \nonumber \\ 
a_k(z) &=& \overline a_{k-1}(z) * \widetilde q_k(z), \\
\overline a_k(z) &=& \textrm{GMR}(a_k(z)),
\end{eqnarray}
and similarly for the backward recursion.

Similarly at the variable node, the product (\ref{eqn:varnode}) can be decomposed into a forward and backward recursion.  In this case as well, the Gaussian mixture reduction algorithm is applied after each step of the recursion.

In the Gaussian mixture reduction algorithm, it is desirable to repeat step 4 as long as the current reduced Gaussian function $g(z)$ (represented by $\mathcal C$) remains a good approximation of the input function $f(z)$ (represented by $\mathcal L$).  In practice, it was found that using a ``local'' stopping condition of a threshold on the one-step error was sufficient to give a good ``global'' approximation $f(z) \approx g(z)$.  In many cases, $f(z)$ was well-approximated by a single Gaussian, which was found by the proposed algorithm.

However, using an error threshold alone does not always restrict the number of output Gaussians, an important goal of the mixture reduction algorithm.   Thus, a second stopping condition, which requires that the number of Gaussians be lower than some fixed threshold, is also enforced.  Thus, the Gaussian combining may continue while $M^{\mathsf c} > M$, even if the one-step error threshold has been exceeded.   In practice, this did not appear to have a detrimental result for a wide range of symbol-error rates.

\begin{figure*}[t]
\begin{center}
\includegraphics[width=9.5cm]{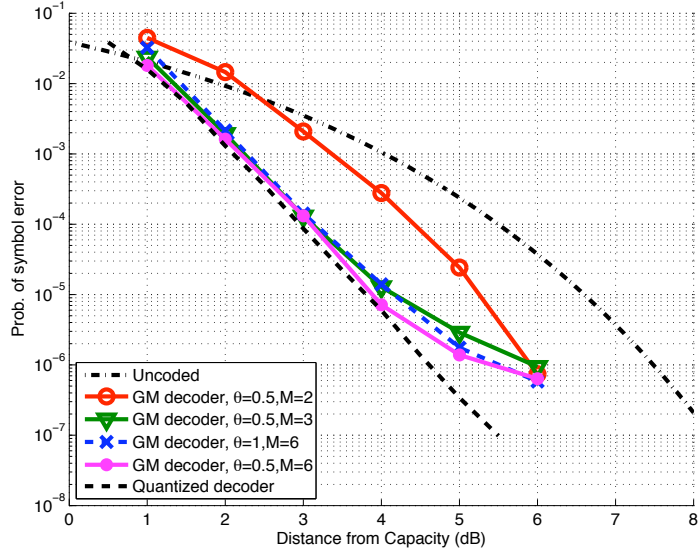}
\end{center}
\caption{Symbol error rate of proposed Gaussian mixture (GM) decoder vs. quantized decoder for $n=100$, $d=5$.}
\label{fig1}
\end{figure*}

Simulation results comparing the proposed decoder with the quantized decoder \cite{Sommer-itsub} are shown in Fig. \ref{fig1}.  A LDLC with $n=100, d=5$ was used.  The symbol error rate of a cubic lattice used for transmission is labeled ``Uncoded.''  The horizontal axis is the difference between the channel noise variance and the Poltyrev capacity, $1/ 2 \pi e$, in dB.

For the parameter selection $\theta = 0.5, 1.0$, and $M \leq 6$, it was found that the proposed algorithm performed with a slight performance loss when the probability of symbol error was greater than $10^{-5}$. For example, with $\theta=0.5$ and $M=6$, the loss at a symbol error rate of $10^{-5}$ is less than 0.1 dB.  For lower symbol error rates, an error floor appears.  It may be helpful to consider this error floor as analogous to quantization error floors which appear in the decoding of low-density parity check codes when insufficient quantization levels are used.

\emph{Complexity}     In the Gaussian mixture reduction algorithm, the primary complexity is computing the initial error, which requires computing the GQL between $N$ pairs, a complexity of $O(N^2)$.  In the Gaussian mixture decoder, the primary complexity the pairwise-computation of the outputs, which is $O(M^2)$.   These numbers $N$ and $M$ are random variables which depend upon the nature of the messages, and the effectiveness of the Gaussian mixture reduction algorithm.  In the simulations the maximum value of $M$ was 6, and $N \leq k M^2$, where $k$ is the constant number of integer shifts, $k=3$ was used in the simulations.  On the other hand, the complexity of the quantized algorithm is dominated by a discrete Fourier transform of size $1/\Delta$ where $\Delta$ is the quantization bin width, $\Delta = 1/128$ was used in the simulations.  It is difficult to directly make comparisons of the computational complexity of the two algorithms.

The memory required for the proposed algorithm, however, is significantly superior.  The proposed algorithm requires storage of $3M$ (for the mean, variance and mixing coefficient), for each message, where $M \leq 6$.  The quantized algorithm, however used 1024 quantization points for each message.

\section{Conclusion \label{sec:conclusion} }

LDLC codes can be used for communication over unconstrained power channels.  In this paper, we  proposed a new LDLC decoding algorithm which exploits the Gaussian nature of the decoder messages.  The core of the algorithm is a Gaussian mixture reduction method, which approximates a message by a smaller number of Gaussians.    As a result, the LDLC algorithm which tracks the means, variances and mixing coefficients of the component Gaussians, rather than using quantized messages, was tractable.  It was shown by computer simulation that this algorithm performs nearly as well as the quantized algorithm, when the dimension is $n=100$, and the probability of symbol error is greater than $10^{-5}$.


\end{document}